\documentclass[aps,twocolumn,notitlepage,superscriptaddress, amsmath,amssymb,amsfonts,floatfix]{revtex4-1}
\usepackage{amsmath}
\usepackage{graphicx}
\usepackage{epsfig}
\usepackage{amssymb}

\usepackage{color}
\usepackage{psfrag}
\usepackage[normalem]{ulem}
\usepackage{float}

\newcommand{\be}{\begin{equation}}
\newcommand{\ee}{\end{equation}}

\begin{document}

\title{Emergent excitability in adaptive networks of non-excitable units}

\author{Marzena Ciszak}
	\affiliation{CNR - Consiglio Nazionale delle Ricerche - Istituto Nazionale di Ottica, Via Sansone 1, I-50019 Sesto Fiorentino (FI), Italy}
	\author{Francesco Marino}
	\affiliation{CNR - Consiglio Nazionale delle Ricerche - Istituto Nazionale di Ottica, Via Sansone 1, I-50019 Sesto Fiorentino (FI), Italy}
	\affiliation{INFN, Sezione di Firenze, Via Sansone 1, I-50019 Sesto Fiorentino (FI), Italy}	
	\author{Alessandro Torcini}
	\affiliation{Laboratoire de Physique Th\'eorique et Mod\'elisation, Universit\'e de Cergy-Pontoise,CNRS, UMR 8089, 95302 Cergy-Pontoise cedex, France}
	\affiliation{CNR - Consiglio Nazionale delle Ricerche - Istituto dei Sistemi Complessi, via Madonna del Piano 10, I-50019 Sesto Fiorentino, Italy}
	\author{Simona Olmi}
	\email[corresponding author: ]{simona.olmi@fi.isc.cnr.it}
	\affiliation{Inria Sophia Antipolis M\'{e}diterran\'{e}e Research Centre, 2004 Route des Lucioles, 06902 Valbonne, France}
	\affiliation{CNR - Consiglio Nazionale delle Ricerche - Istituto dei Sistemi Complessi, via Madonna del Piano 10, I-50019 Sesto Fiorentino, Italy}	 
	\affiliation{INFN, Sezione di Firenze, Via Sansone 1, I-50019 Sesto Fiorentino (FI), Italy}	
	\date{\today}
	
\begin{abstract}
Population bursts in a large ensemble of coupled elements result from the interplay between the local excitable properties of the nodes and the global network topology. Here collective excitability and self-sustained bursting oscillations are shown to spontaneously emerge in adaptive networks of globally coupled non-excitable units. The ingredients to observe collective excitability are the coexistence of states with different degree of synchronizaton joined to a global feedback acting, on a slow timescale, against the synchronization (desynchronization) of the oscillators. These regimes are illustrated for two paradigmatic classes of coupled rotators: namely, the Kuramoto model with and without inertia. For the bimodal Kuramoto model we analytically show that the macroscopic evolution originates from the existence of a critical manifold organizing the fast collective dynamics on a slow timescale. Our results provide evidence that adaptation can induce excitability by maintaining a network permanently out-of-equilibrium.
\end{abstract}

\date{\today}

\maketitle

\paragraph*{Introduction}
Complex networks composed of simple elements, usually rotators, have been widely analyzed in the last decades in order to identify the emergence of non-trivial macroscopic phenomena, ranging from synchronization to 
collective oscillations, quasiperiodicity and chaos \cite{kuramoto2003, pikovsky2003, acebron2005, matthews1990, hakim1992, nakagawa1993}. The field has particularly fluorished in the last years thanks to the development of analytic techniques to obtain exact low dimensional mean-field descriptions for phase oscillators networks \cite{ott2008}. 

Despite this intense activity only a few analyses have reported signatures of collective excitable features in such networks \cite{so2011,skardal2014}. Excitable systems appear in many fields of science and are particularly studied in the context of mathematical neuroscience as simplified descriptions of neural systems \cite{tuckwell1988,kock1999}. They are characterized by a (quiescent) state that is linearly stable, but susceptible to finite-amplitude perturbations. The return to equilibrium entails a large excursion in the phase space corresponding to the emission of a pulse of well-defined amplitude and duration. The reinjection mechanism to the excitable quiescent state is often related to the competition of multiple time-scales \cite{izhikevich2000}. The dynamical scenario emerging in these low dimensional slow-fast systems can be extremely rich displaying regular as well as chaotic spiking and bursting behaviours joined to complex bifurcation structures \cite{hindmarsh1984,terman1992,wang1993,mosekilde2001,gonzalez2003,innocenti2007}.

Collective excitable responses and bursting activities have been previously reported \cite{coombes}, e.g. in diffusively-coupled spatially-extended systems where they appear in the form of excitable waves \cite{meron_rev}, as transient synchronization states in arrays of coupled units \cite{mchaos09,dolcem}. In all these cases however, the dynamics fully relies on excitable features of the nodes.
 
In this Rapid Communication we show that a self-sustained adaptation mechanism can give rise to an out-of-equilibrium scenario, where a network of non-excitable nodes, permanently driven across a hysteretic phase-transition, can become collectively excitable.
In particular, we investigate the effects of a global linear feedback on the dynamics of the Kuramoto model (KM) with and without inertia. In absence of adaptation, the considered networks display hysteretic first-order transitions involving asynchronous (AS) and partially synchronized (PS) states, as well as standing waves \cite{SW,pazo2009,tanaka1997,olmi2016}. In presence of the feedback, these systems reveal collective dynamical features typical of excitable models, despite a non-excitable single node dynamics. The origin of these behaviours is related to the competition of
the fast synchronization/desynchronization phenomena triggered by the slow adaptation. For the bimodal KM, we derive an exact three dimensional slow-fast mean-field formulation, which allows to interpret all the observed collective regimes in terms of an attractive invariant manifold, on which the (slow) mean-field dynamics takes place.
Finally, these phenomena are shown to emerge also in the Kuramoto model with inertia (KMI),
confirming the generality of our results.

\paragraph*{The model} We consider a globally coupled network of $N$ rotators with adaptive coupling strength $S(t)$, which reads as
\begin{subequations}
\label{network} 
\begin{eqnarray}
m {\ddot \theta}_i + {\dot \theta}_i(t) &=& \omega_i + \frac{S(t)}{N} \sum_{j=1}^N \sin(\theta_j(t)- \theta_i(t)) \;  \label{net1} \\
{\dot S}(t) &=& \varepsilon \left[-S(t) + K -\alpha R(t) \right] \;  \label{net2}
\end{eqnarray}
\end{subequations}
where $\theta_i$ ($\omega_i$) are the phases (natural frequencies) of each rotator and $m$ their mass.
As stated in Eq. \eqref{net2}, the evolution of $S(t)$ is controlled, via a linear feedback \cite{skardal2014}, by 
$R(t)$, which is the modulus of the complex Kuramoto order parameter 
$Z(t) = \frac{1}{N} \sum_{j=1}^N {\rm e}^{i \theta_j(t)} = R(t) {\rm e}^{i \phi(t)}$ \cite{kuramoto}.
The macroscopic variable $R$ measures the level of synchronization among the rotators: AS (PS) dynamics will correspond to $R = 0$ ($0 < R \le 1$). The gain of the feedback loop is controlled by $\alpha$ and its bandwidth by $\varepsilon$, therefore, depending on the value of $R(t)$, the coupling $S(t)$ 
can range between $K -\alpha$ ($R=1$) to $K$ ($R=0$). 
We assume $0 < \varepsilon \ll 1$, i.e. the modulation of the coupling is 
slow with respect to the switching times between incoherent and coherent states.

In absence of feedback ($\alpha=0$), the equation (\ref{net1}) reduces to the KMI
with coupling constant $S(t) \equiv K$ \cite{tanaka1997} and for $m=0$ to the standard KM \cite{kuramoto}.
For both these models, at sufficiently low (large) coupling, one has a desynchronized (partially synchronized) regime. 
If the transition from AS to PS dynamics is continuous, as for the KM with
unimodal frequency distribution, the feedback \eqref{net2} has only the effect to renormalize the
coupling strength, but the collective dynamics will always converge to a stable fixed point for all the 
parameter values. This is no more the case if the uncontrolled system displays a first-order hysteretic transition
from incoherence to coherence, as it occurs for the KMI  \cite{tanaka1997,olmi2014,olmi2016}
and for the KM with a bimodal frequency distribution \cite{pazo2009}.
In this case, the linear feedback introduced above can give rise to a wealth of macroscopic behaviours over multiple timescales, including excitability and periodic/chaotic spiking and bursting oscillations.

\paragraph*{KM with bimodal frequency distribution} 

\begin{figure}
\begin{center}
\includegraphics[width=0.9 \linewidth]{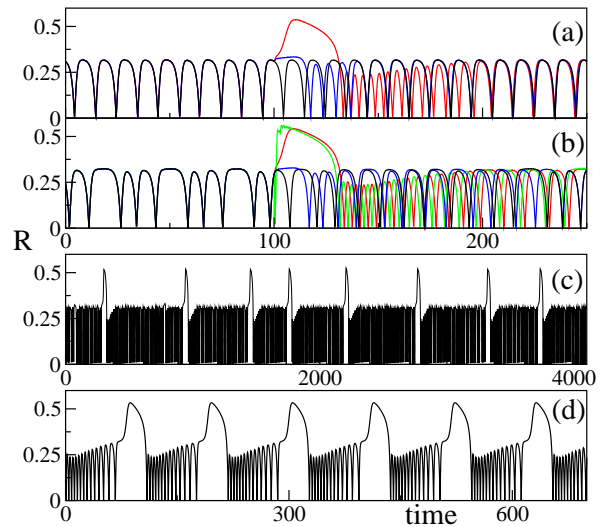}
\end{center}
\caption{Synchronization parameter $R$ versus time for the bimodal Kuramoto network for different dynamical regimes: (a) periodic spiking ($K=7$); (b) chaotic spiking ($K=7.403$); chaotic bursting ($K=7.405$); periodic bursting ($K=7.41$). Panels (a) and (b) display also the system response to perturbations of different amplitudes $A$ of the collective variable $S(t)$ : subthreshold responses for $A=0.10255$ (blue traces);
excitable responses for $A=0.15$ (red traces) and $A=0.1575$ (green trace). Other parameters: 
$\epsilon=0.01$,  $\Delta=1.4$, $\omega_0=1.8$, $\alpha=5$, and network size $N=500,000$.}
\label{fig1}
\end{figure}

As a first paradigmatic example of networks displaying hysteretic phase-transitions, we consider 
the KM  with a bimodal distribution of natural frequencies \cite{pazo2009}. In particular, 
in order to be able to derive an exact mean-field description of the model we consider
a bimodal distribution  given by the sum of two Lorentzians centered at $\pm \omega_0$ and with 
half-width at half-maximum $\Delta$ \cite{bimodal}. For the chosen parameters ($\omega_0=1.8$
and $\Delta=1.4$), in absence of feedback, we observe a coexistence regime between travelling waves and PS states \cite{martens2009}.
By fixing a finite feedback gain $\alpha$ and the frequency cutoff $\varepsilon$ and increasing the control parameter $K$, we observe the sequence of macroscopic regimes displayed in Fig. \ref{fig1} in terms of the synchronization parameter $R(t)$. 

At small $K$ values the network is essentially desynchronized, apart finite size fluctuations
associated with $R \simeq {\cal O} (1/\sqrt{N})$. For $K$ larger than a critical value, one observes 
the emergence of periodic collective oscillations, alternating PS phases with abrupt desynchronization events ({\it spikes}), as shown in Fig. \ref{fig1} (a).
Further increasing $K$ leads first to an increase of the interspike period and then to a chaotic phase (see Fig. \ref{fig1} (b)). In the chaotic and periodic spiking regimes the system is excitable: small perturbations of the collective variable $S(t)$ elicit rapidly decaying responses in $R(t)$ (blue traces), while sufficiently strong stimuli induce a large degree of synchronization, corresponding to a {\it burst} with a well-defined shape, amplitude and duration (red traces in Figs. \ref{fig1} (a) and (b)).
In Fig. \ref{fig1} (b) we also show that, except for the initial rise-time, the burst orbit is barely affected by higher-amplitude perturbations (green trace), thus confirming an important feature of excitable systems. The chaotic nature of the spiking dynamics can be appreciated in Fig. \ref{fig1} (b), where in response to pertubations of different amplitude, the system relaxes to different final trajectories, in contrast to what shown in Fig. \ref{fig1} (a) for a periodic spiking regime. By further increasing the
parameter $K$ one observes the emergence of a regime characterized by the presence of bursts separated
by many spikes, whose number appears to be irregular (Fig. \ref{fig1} (c)). This bursting phase is chaotic, as
we will verify in the following. As shown in Fig. \ref{fig1} (d), a further increase of $K$ leads to a periodic bursting state, where the bursts are separated by a fixed number of spikes.  The number of spikes decreases for growing $K$ and finally we observe a stationary regime characterized by a finite value of $R$ for sufficiently large $K$. Similar spiking and bursting regimes, as well as period adding-sequences, are typical for low-dimensional slow-fast systems possessing some attractive manifolds on which the dynamics evolves slowly. In this context, a paradigmatic example is represented by the Hindmarsh-Rose neuronal model \cite{hindmarsh1984,innocenti2007}. Similarly, in our network, collective excitability and bursting phenomena originate from the existence of a critical manifold which organizes the mean-field dynamics on the slow time scale, as we will show in the next paragraph.
   
\paragraph*{Exact mean-field analysis} 
To better understand the observed phenomenology we derive an exact mean-field dynamics for the network \eqref{network}, by extending the macroscopic formulation derived in \cite{martens2009,so2011}
for the bimodal KM, based on the Ott-Antonsen Ansatz \cite{ott2008}.
In particular, by following \cite{martens2009}, one can rewrite the complex order parameter $Z$
in terms of two sub-population order parameters $z_k=\rho_k{\rm e}^{i \phi_k}$ ($k=1,2$),
each relative to a Lorentzian distribution, as $Z = \frac{1}{2}(z_1 + z_2)$. Moreover, by assuming $\rho_1 \approx \rho_2 = \rho$, one arrives
to the following equations ruling the macroscopic evolution of the network
\begin{subequations}
\label{mf} 
\begin{eqnarray}
\dot{\rho} & = & -\Delta \rho + \frac{S}{4} \rho (1 - \rho^2) (1 + \cos(\phi)) \;  \label{eq:model1} \\
\dot{\phi} & = & 2 \omega_0 - \frac{S}{2} (1 + \rho^2) \sin(\phi) \;  \label{eq:model2} \\
\dot{S} & = & -\epsilon \left[ S - K + \alpha \rho   \sqrt{\frac{1 + \cos(\phi)}{2}} \right] \;  \label{eq:model3} 
\end{eqnarray}
\end{subequations}
where $\phi = \phi_2 - \phi_1$ and
the global feedback equation (\ref{eq:model3}) directly follows from (\ref{net2}) by noticing that
$R \equiv \rho \sqrt{(1 + \cos(\phi))/2}$.

\begin{figure}
\centering
\includegraphics[width=1\linewidth]{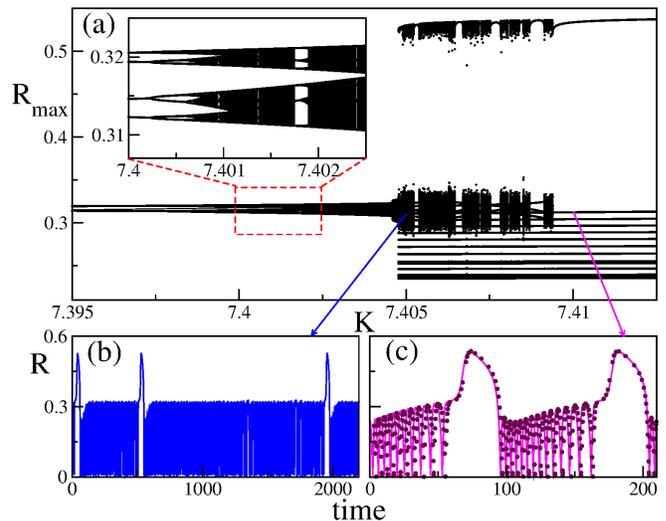}
\caption{(a) Bifurcation diagram of the mean-field model \eqref{mf}. The maxima of $R$ are displayed as a function of the parameter $K$. Time-traces of $R$ for the mean-field model are shown in (b) for the chaotic bursting regime ($K=7.405$) and in (c) 
for the periodic bursting phase ($K=7.41$). Symbols in panel (c) are the $R$ values estimated by the direct simulation of the network \eqref{network} of size $N=500,000$.
Other parameters: $\epsilon=0.01$, $\omega_0=1.8$, $\Delta=1.4$ , $\alpha=5$.}
	\label{fig2}
\end{figure}

The dynamics of the mean-field model \eqref{mf} is attracted towards a stable fixed point corresponding to
an AS regime ($R=0$) for $K < K_H = 4 \Delta$, while at $K_H$ a supercritical Hopf bifurcation takes place giving rise to a stable limit cycle (periodic spiking). As shown in Fig. \ref{fig2}(a), the Hopf bifurcation is followed by a period doubling cascade leading to a chaotic spiking regime for $K > 7.401$. Both the regular and chaotic spiking oscillators display excitable features analogous to the collective ones displayed in Fig. \ref{fig1} (a) and (b) for the corresponding network. At $K_c \simeq 7.40477$, we observe an abrupt increase of the size of the attractor for $R$, that corresponds to the appearence of chaotic bursts. This transition resembles a crisis that destabilizes the chaotic attractor associated to the spiking phase, leading to the emergence 
of a new chaotic attractor characterized by bursting dynamics, as shown in Fig. \ref{fig2}(b). Similar transitions
have been reported for single excitable systems in \cite{wang1993,gonzalez2003}.
Moreover, we have measured the maximal Lyapunov exponent $\lambda_m$ \cite{pikovsky2016} for the mean-field model \eqref{mf} and observed that this is positive for $K \in [7.401,7.408]$, a region containing the transition point $K_c$. This demonstrates that we can have collective chaos \cite{hakim1992,nakagawa1993,shibata1998,olmi2011} also for a single Kuramoto population with self-generated collective adaptation and not only in presence of an external periodic forcing \cite{so2011} or for multiple coupled populations \cite{bick2018}.
Furthermore, $\lambda_m$ shows a pronunciated peak in correspondence of $K \simeq K_c$, similarly to what reported in \cite{innocenti2007} for the single Hindmarsh-Rose model.
For $ K > 7.408$ we observe periodic bursting attractors, as the one reported in Fig. \ref{fig2}(c) for the mean-field model. In Fig. \ref{fig2}(c) are also reported the data obtained from the network, for the macroscopic indicator $R$: the excellent agreement confirms the validity of the mean-field reduction. For larger $K$ we observe a decrease in the number of spikes separating
two successive periodic bursts up to $K=8.15843$ when the system returns to a stable fixed point with $R > 0$
via a sub-critical Hopf bifurcation.

\begin{figure}
\includegraphics[width=1\linewidth]{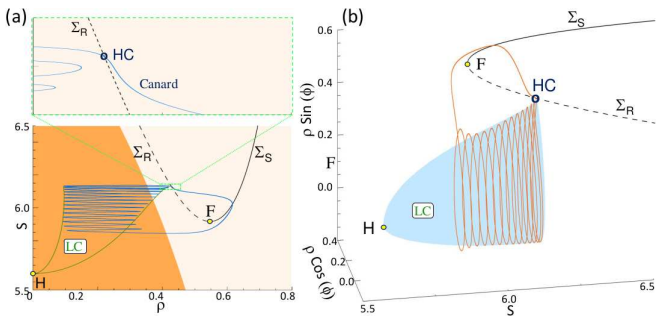}
\caption{Critical manifold of the mean-field equations (\ref{eq:model1}-\ref{eq:model3}) together with a bursting solution
(blue solid line). (a) Projection on the ($\rho$ $S$) plane: solid and dashed black lines indicate the attracting ($\Sigma_S$)
repelling ($\Sigma_R$) manifolds, separated by the fold point $F$. In the orange shaded area $\dot{S}>0$. The green curves denote the extrema of the limit cycle ($LC$) emerging from the Hopf bifurcation $H$. Inset: enlarged view of the transition region from the spiking regime towards $\Sigma_S$ occuring via a saddle homoclinic orbit bifurcation ($HC$). (b) Three-dimensional representation in the space ($S$,$\rho \cos(\phi)$, $\rho \sin(\phi)$) disclosing the bursting regime. Parameters are $\omega_0=1.8$, $\Delta=1.4$, $\alpha=5.0$, $\epsilon=0.01$, $K=7.41$.}
\label{fig3}
\end{figure}

\paragraph*{Geometric singular perturbation analysis of the mean-field model}

Since $\epsilon \ll 1$, the adaptive variable $S$ evolves at a much slower rate than $\phi$ and $\rho$. Hence the dynamics of Eqs. \eqref{mf} splits into periods of fast and slow motion that can be analyzed separately \cite{hirsch1974}. 
On the fast time scale $t$, the evolution is described by the mean-field equations (\ref{eq:model1}-\ref{eq:model2}) (fast sub-system) with $S$ acting as a slowly varying adiabatic parameter. The equilibria of this dynamical sub-system lay on the one-dimensional manifold $\Sigma= \Sigma_0 \cup \Sigma_{\rho}$, where $\Sigma_0$ is given by the set of incoherent steady-state solutions $\Sigma_0 = \{\rho_s = 0, \sin(\phi_s) = 4 \omega_0 /S, S\}$, and $\Sigma_{\rho}= \{\rho_s, \phi_s, S\}$ is defined by the equations $S (1+\rho_s^2) \sin(\phi_s)=4 \omega_0$ and 
\begin{equation}
S =\frac{2\omega_0^2}{\Delta} \frac{1-\rho^2}{(1+\rho^2)^2} + \frac{2 \Delta }{1-\rho^2} \equiv \mathcal{F}(\rho).
\label{eq_cm}
\end{equation}
On the slow time scale $\tau=\epsilon t$, the motion is governed by the feedback equation Eq. (\ref{eq:model3}) with the algebraic constraint $(\dot{\rho},\dot{\phi})$=($0,0$). The fixed points of the fast sub-system thus define the critical manifold on which the slow dynamics take place. Since the trajectories of Eqs. (\ref{eq:model1}-\ref{eq:model3}) will be attracted by the stable parts of $\Sigma$, while will be repelled by the unstable ones \cite{fenichel1979}, the stability properties of the critical manifold determine the dynamics. 
Linearizing the fast sub-system on $\Sigma_{\rho}$, we find that it consists of a branch of stable equilibria $\Sigma_S$ (solid line in Fig. \ref{fig3}) and an unstable one $\Sigma_R$ (dashed line) coalescing in a saddle-node bifurcation at the fold point $F$.

For $\omega_0 > \Delta$, the equilibria along $\Sigma_0$ (i.e. for $S > 4\omega_0$) are always unstable.
At lower values of $S$, the fast sub-system (\ref{eq:model1}-\ref{eq:model2}) displays a supercritical Hopf bifurcation at $S_H=K_H$, leading to the emergence of a stable limit cycle with $\rho >0$.


The critical manifold is thus composed by an attracting part $\Sigma_S$ and two repelling parts $\Sigma_R$ and $\Sigma_0$.
Moreover, for some values of the slow variable, the above stationary states coexist with a multiplicity of stable limit cycles emanating from the Hopf bifurcation at $K_H$. 
On this basis, we can explain the appearence of bursting in our system.
In Fig. \ref{fig3}a we plot the projection of the critical manifold on the $(\rho, S)$ plane together with a solution of Eqs. (\ref{eq:model1}-\ref{eq:model3}) in the bursting regime (blue solid line).

Starting, e.g., from a PS initial condition, the motion is rapidly attracted by $\Sigma_S$.
Since Eq. (\ref{eq:model3}) dictates that $S$ is always decreasing on the curve $\mathcal{F}(\rho)$ (see \ref{fig3}a), the system is driven towards the fold point $F$, where it forcibly leaves the critical manifold turning on the fast dynamics transversal to it. When the trajectory enters the region in which the fast sub-system has a stable limit cycle (green lines Fig. \ref{fig3}a and shaded area in \ref{fig3}b), the slow monotonic evolution translates into a sequence of nearly-periodic spikes. 
Such an oscillatory state persists until it collides with the repelling part of the manifold where it disappears via a saddle homoclinic-orbit bifurcation ($HC$). Since $\Sigma_{R}$ repels all neighboring trajectories, while $\Sigma_{S}$ attract them, the motion is driven back to the upper state where a new bursting cycle begins.
\begin{figure}
\begin{center}
\includegraphics[width=1.\linewidth]{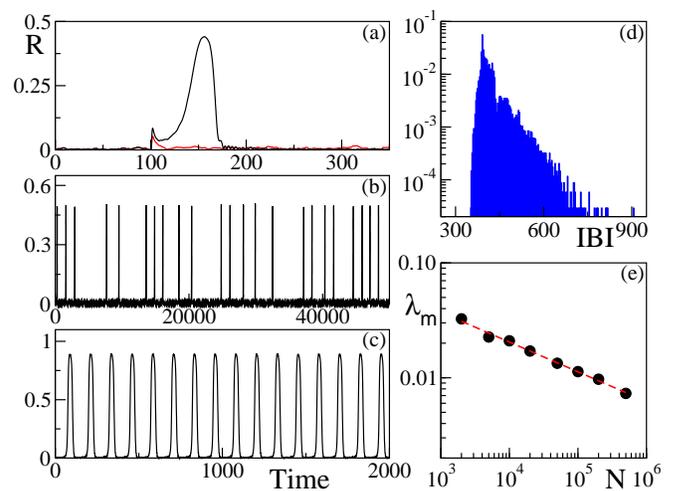}
\end{center}
\caption{Time-series of $R(t)$ for the unimodal KMI: (a) $K=4.5$ (excitable response); (b) $K=4.75$ (irregular bursting); (c) $K=10.3$ (periodic bursting). For $K=4.75$ we report also the IBI distribution (d) and the maximal Lyapunov exponent versus $N$ (e). The red dashed line in (e) refers to a power law fitting with exponent $\simeq 0.259$. Other parameters: $N=500000$, $m=2$,  $\epsilon=0.01$, $\alpha=30$.}
\label{fig4}
\end{figure}	
Interestingly, the transition from the spiking activity to the upper-state exhibits the typical features of canard explosions \cite{canard1,canard2}: due to the finiteness of $\varepsilon$, the trajectories close to $\Sigma_{R}$ do not jump immediately to $\Sigma_{S}$, but continue moving on the slow time-scale along the unstable portion of the manifold for a certain amount of time (see inset in Fig. \ref{fig3}a)). A more detailed analysis can be found in the Supplemental Material \cite{suppl}.

The above scenario
is known as \emph{fold-homoclinic} bursting \cite{izhikevich2000}: it has been found in several low dimensional neuron models, including e.g. the Hindmarsh-Rose \cite{hindmarsh1984} and the Morris–Lecar system with current-feedback control \cite{izhikevich2000}. 

Excitability and chaotic bursting can be explained on the same basis: when the system is in the regime of quasi-harmonic oscillations, either periodic or chaotic, an external perturbation or a sufficiently large chaotic fluctuation can trigger the fast dynamics, giving rise to an excursion to the upper stable branch $\Sigma_S$ before returning to initial state \cite{marino2007,al2009,al2010}.

\paragraph*{Dynamics of the KMI network}
To further support the generality of the phenomenon, we now consider a KMI
with mass $m=2$ and natural frequencies distributed according to either an unimodal 
or a bimodal Gaussian distribution.
  
Typical time-traces of $R(t)$ for the KMI with unimodal distribution centered in zero and with unitary standard deviation,
are plotted in Fig. \ref{fig4} for different values of $K$. In this case we observe collective excitability and 
{\it fold-fold} bursting \cite{izhikevich2000}, but there is no trace of spiking dynamics. This is probably due to the fact 
that the fast sub-system displays a hysteretic transition from an AS to a PS regime, but no evidences of 
standing waves as in the previous case. We have found irregular busting at low $K$ ($ 4.6 \lesssim K \lesssim 7$),
as shown in Fig. \ref{fig4} (b), and periodic bursting for sufficienly large $K \gtrsim 7 $, see Fig. \ref{fig4} (c). 
The erraticity in the bursting dynamics is confirmed by the distribution of the Inter-Burst Intervals (IBIs) among successive bursts, which displays a clear exponential tail charateristic of Poissonian processes. The origin of the irregular bursting is related to a weak form of chaos induced by finite-size fluctuations \cite{popovych2005}. Indeed the maximal Lyapunov exponent vanishes in the thermodynamic limit as $\lambda_m \propto N^{-1/4}$,  as shown in Fig. \ref{fig4} (d). 

\begin{figure}
\begin{center}
\includegraphics[width=1.\linewidth]{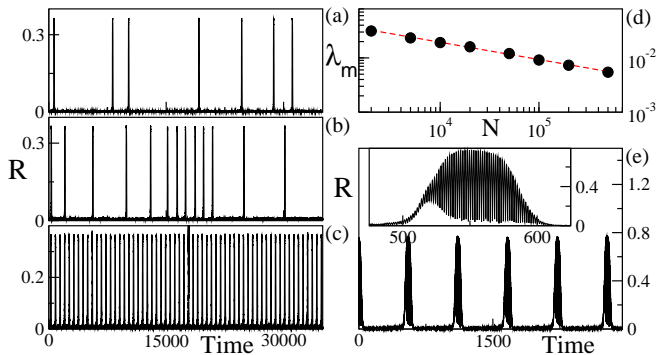}
\end{center}
\caption{Time-series of $R(t)$ for the bimodal KMI. 
Erratic busting behaviour is reported in (a-c) at $K=9.45$ for increasing system sizes: $N=100000$ (a),
200000 (b) and 500000 (c). The corresponding maximal Lyapunov exponent $\lambda_m$ is reported in (d) versus $N$,
the dashed red line refers to a power-law fitting with exponent $\simeq 0.318$.
In (e), $R(t)$ is displayed versus time for $K=10.3$ and $N=200000$;
an enlargement is shown in the inset.
Other parameters: $m=2$, $\epsilon=0.01$, $\alpha=30$.}
\label{fig5}
\end{figure}

As a final example we consider a KMI network with a bimodal distribution composed by two almost non overlapping Gaussians 
centered at  $\omega_0=\pm 2$ and with unitary standard deviations \cite{olmi2016}. In this case we still observe
regimes of irregular bursting due to finite size effects ($8.5 \lesssim K \lesssim 10$). 
Indeed, as shown in Fig. \ref{fig5} (a-c), the bursts
become rarer for increasing $N$, while the maximal Lyapunov exponent vanishes as $\lambda_m \propto N^{-1/3}$ (see
Fig. \ref{fig5} (d)). Besides this regime, for larger $K$, we can also have periodic bursting. However, 
the oscillations now emerge on the top of the burst, as shown in Fig. \ref{fig5} (e), due to the coexistence of
standing waves with the PS regime for the fast sub-system as found in \cite{olmi2016}.
These bursts resemble the so called {\it Hopf-Hopf} bursting reported in \cite{izhikevich2000}. Similar bursting is also observed in the unimodal KMI for large masses.

\paragraph*{Conclusions}  
 
Our analysis indicate the minimal ingredients for the emergence of collective excitability and bursting oscillations in adaptive networks of rotators. The network dynamics without feedback, corresponding to the fast sub-system, must display a hysteretic phase transition connecting a low synchronization state to one with a higher synchronization degree. The global feedback equation (slow sub-system) introduces a state-dependent modulation of a control parameter (in our case, the coupling strength), driving the system across the hysteresis cycle.

For the bimodal KM, an exact mean-field formulation can be derived consisting of a 
3D system with two fast and one slow variable  .
A detailed geometric singular perturbation analysis of this model allows us 
to explain collective excitability in terms of the stability properties of a 1D
slow invariant manifold \cite{suppl}. This demonstrates that the phenomenon persists
in the thermodynamic limit and it is not related to finite-size
effects \cite{skardal2014}.
Furthermore, for this autonomous model we have shown the existence of new types of collective chaos
(namely, chaotic spiking and bursting) and characterized in details the transition between the two
chaotic regimes. The reported dynamical macroscopic scenario is pretty reminiscent of that observed for the 
Hindmarsh-Rose model for a single neuron \cite{wang1993,gonzalez2003,innocenti2007}. This analogy paves the way for the application of our results in the context of computational neurosciences.

For what concerns the KMI, we observe only bursting dominated phases, which can be weakly chaotic, but in the
thermodynamic limit we expect regular trains of fold-fold or Hopf-Hopf bursts \cite{izhikevich2000} only. 
As shown in previous analysis \cite{olmi2016}, for sufficiently large masses, the KMI (in absence of
feedback) can become chaotic, therefore for $m \gg 1$, we expect chaotic bursting regimes to appear also for this network.

The exact reduction techniques developed for networks of phase oscillators \cite{ott2008,martens2009}
do not apply to the KMI. In the latter case, a promising approach to investigate, in order to 
derive a low dimensional mean-field description, is 
the so-called circular cumulant expansion, recently applied with success to noisy oscillator populations \cite{tyulkina2018}
and to stochastic systems with small inertia \cite{goldobin2020}.

\begin{acknowledgments}
A.T. received financial support by the Excellence Initiative I-Site Paris Seine (Grant No. ANR-16-IDEX-008), by the Labex MME-DII (Grant No ANR-11-LBX-0023-01), and by the ANR Project ERMUNDY (Grant No ANR-18-CE37-0014), all part of the French program Investissements d’Avenir. 
\end{acknowledgments}


\begin{thebibliography}{0}%
\makeatletter
\providecommand \@ifxundefined [1]{%
 \@ifx{#1\undefined}
}%
\providecommand \@ifnum [1]{%
 \ifnum #1\expandafter \@firstoftwo
 \else \expandafter \@secondoftwo
 \fi
}%
\providecommand \@ifx [1]{%
 \ifx #1\expandafter \@firstoftwo
 \else \expandafter \@secondoftwo
 \fi
}%
\providecommand \natexlab [1]{#1}%
\providecommand \enquote  [1]{``#1''}%
\providecommand \bibnamefont  [1]{#1}%
\providecommand \bibfnamefont [1]{#1}%
\providecommand \citenamefont [1]{#1}%
\providecommand \href@noop [0]{\@secondoftwo}%
\providecommand \href [0]{\begingroup \@sanitize@url \@href}%
\providecommand \@href[1]{\@@startlink{#1}\@@href}%
\providecommand \@@href[1]{\endgroup#1\@@endlink}%
\providecommand \@sanitize@url [0]{\catcode `\\12\catcode `\$12\catcode
  `\&12\catcode `\#12\catcode `\^12\catcode `\_12\catcode `\%12\relax}%
\providecommand \@@startlink[1]{}%
\providecommand \@@endlink[0]{}%
\providecommand \url  [0]{\begingroup\@sanitize@url \@url }%
\providecommand \@url [1]{\endgroup\@href {#1}{\urlprefix }}%
\providecommand \urlprefix  [0]{URL }%
\providecommand \Eprint [0]{\href }%
\providecommand \doibase [0]{http://dx.doi.org/}%
\providecommand \selectlanguage [0]{\@gobble}%
\providecommand \bibinfo  [0]{\@secondoftwo}%
\providecommand \bibfield  [0]{\@secondoftwo}%
\providecommand \translation [1]{[#1]}%
\providecommand \BibitemOpen [0]{}%
\providecommand \bibitemStop [0]{}%
\providecommand \bibitemNoStop [0]{.\EOS\space}%
\providecommand \EOS [0]{\spacefactor3000\relax}%
\providecommand \BibitemShut  [1]{\csname bibitem#1\endcsname}%
\let\auto@bib@innerbib\@empty
\end{thebibliography}%


\begin{thebibliography}{45}%
\makeatletter
\providecommand \@ifxundefined [1]{%
 \@ifx{#1\undefined}
}%
\providecommand \@ifnum [1]{%
 \ifnum #1\expandafter \@firstoftwo
 \else \expandafter \@secondoftwo
 \fi
}%
\providecommand \@ifx [1]{%
 \ifx #1\expandafter \@firstoftwo
 \else \expandafter \@secondoftwo
 \fi
}%
\providecommand \natexlab [1]{#1}%
\providecommand \enquote  [1]{``#1''}%
\providecommand \bibnamefont  [1]{#1}%
\providecommand \bibfnamefont [1]{#1}%
\providecommand \citenamefont [1]{#1}%
\providecommand \href@noop [0]{\@secondoftwo}%
\providecommand \href [0]{\begingroup \@sanitize@url \@href}%
\providecommand \@href[1]{\@@startlink{#1}\@@href}%
\providecommand \@@href[1]{\endgroup#1\@@endlink}%
\providecommand \@sanitize@url [0]{\catcode `\\12\catcode `\$12\catcode
  `\&12\catcode `\#12\catcode `\^12\catcode `\_12\catcode `\%12\relax}%
\providecommand \@@startlink[1]{}%
\providecommand \@@endlink[0]{}%
\providecommand \url  [0]{\begingroup\@sanitize@url \@url }%
\providecommand \@url [1]{\endgroup\@href {#1}{\urlprefix }}%
\providecommand \urlprefix  [0]{URL }%
\providecommand \Eprint [0]{\href }%
\providecommand \doibase [0]{http://dx.doi.org/}%
\providecommand \selectlanguage [0]{\@gobble}%
\providecommand \bibinfo  [0]{\@secondoftwo}%
\providecommand \bibfield  [0]{\@secondoftwo}%
\providecommand \translation [1]{[#1]}%
\providecommand \BibitemOpen [0]{}%
\providecommand \bibitemStop [0]{}%
\providecommand \bibitemNoStop [0]{.\EOS\space}%
\providecommand \EOS [0]{\spacefactor3000\relax}%
\providecommand \BibitemShut  [1]{\csname bibitem#1\endcsname}%
\let\auto@bib@innerbib\@empty
\bibitem [{\citenamefont {Kuramoto}(2003)}]{kuramoto2003}%
  \BibitemOpen
  \bibfield  {author} {\bibinfo {author} {\bibfnamefont {Y.}~\bibnamefont
  {Kuramoto}},\ }\href@noop {} {\emph {\bibinfo {title} {Chemical oscillations,
  waves, and turbulence}}}\ (\bibinfo  {publisher} {Courier Corporation},\
  \bibinfo {year} {2003})\BibitemShut {NoStop}%
\bibitem [{\citenamefont {Pikovsky}\ \emph {et~al.}(2003)\citenamefont
  {Pikovsky}, \citenamefont {Kurths}, \citenamefont {Rosenblum},\ and\
  \citenamefont {Kurths}}]{pikovsky2003}%
  \BibitemOpen
  \bibfield  {author} {\bibinfo {author} {\bibfnamefont {A.}~\bibnamefont
  {Pikovsky}}, \bibinfo {author} {\bibfnamefont {J.}~\bibnamefont {Kurths}},
  \bibinfo {author} {\bibfnamefont {M.}~\bibnamefont {Rosenblum}}, \ and\
  \bibinfo {author} {\bibfnamefont {J.}~\bibnamefont {Kurths}},\ }\href@noop {}
  {\emph {\bibinfo {title} {Synchronization: a universal concept in nonlinear
  sciences}}},\ Vol.~\bibinfo {volume} {12}\ (\bibinfo  {publisher} {Cambridge
  university press},\ \bibinfo {year} {2003})\BibitemShut {NoStop}%
\bibitem [{\citenamefont {Acebr{\'o}n}\ \emph {et~al.}(2005)\citenamefont
  {Acebr{\'o}n}, \citenamefont {Bonilla}, \citenamefont {Vicente},
  \citenamefont {Ritort},\ and\ \citenamefont {Spigler}}]{acebron2005}%
  \BibitemOpen
  \bibfield  {author} {\bibinfo {author} {\bibfnamefont {J.~A.}\ \bibnamefont
  {Acebr{\'o}n}}, \bibinfo {author} {\bibfnamefont {L.~L.}\ \bibnamefont
  {Bonilla}}, \bibinfo {author} {\bibfnamefont {C.~J.~P.}\ \bibnamefont
  {Vicente}}, \bibinfo {author} {\bibfnamefont {F.}~\bibnamefont {Ritort}}, \
  and\ \bibinfo {author} {\bibfnamefont {R.}~\bibnamefont {Spigler}},\
  }\href@noop {} {\bibfield  {journal} {\bibinfo  {journal} {Reviews of modern
  physics}\ }\textbf {\bibinfo {volume} {77}},\ \bibinfo {pages} {137}
  (\bibinfo {year} {2005})}\BibitemShut {NoStop}%
\bibitem [{\citenamefont {Matthews}\ and\ \citenamefont
  {Strogatz}(1990)}]{matthews1990}%
  \BibitemOpen
  \bibfield  {author} {\bibinfo {author} {\bibfnamefont {P.~C.}\ \bibnamefont
  {Matthews}}\ and\ \bibinfo {author} {\bibfnamefont {S.~H.}\ \bibnamefont
  {Strogatz}},\ }\href@noop {} {\bibfield  {journal} {\bibinfo  {journal}
  {Physical review letters}\ }\textbf {\bibinfo {volume} {65}},\ \bibinfo
  {pages} {1701} (\bibinfo {year} {1990})}\BibitemShut {NoStop}%
\bibitem [{\citenamefont {Hakim}\ and\ \citenamefont
  {Rappel}(1992)}]{hakim1992}%
  \BibitemOpen
  \bibfield  {author} {\bibinfo {author} {\bibfnamefont {V.}~\bibnamefont
  {Hakim}}\ and\ \bibinfo {author} {\bibfnamefont {W.-J.}\ \bibnamefont
  {Rappel}},\ }\href@noop {} {\bibfield  {journal} {\bibinfo  {journal}
  {Physical Review A}\ }\textbf {\bibinfo {volume} {46}},\ \bibinfo {pages}
  {R7347} (\bibinfo {year} {1992})}\BibitemShut {NoStop}%
\bibitem [{\citenamefont {Nakagawa}\ and\ \citenamefont
  {Kuramoto}(1993)}]{nakagawa1993}%
  \BibitemOpen
  \bibfield  {author} {\bibinfo {author} {\bibfnamefont {N.}~\bibnamefont
  {Nakagawa}}\ and\ \bibinfo {author} {\bibfnamefont {Y.}~\bibnamefont
  {Kuramoto}},\ }\href@noop {} {\bibfield  {journal} {\bibinfo  {journal}
  {Progress of Theoretical Physics}\ }\textbf {\bibinfo {volume} {89}},\
  \bibinfo {pages} {313} (\bibinfo {year} {1993})}\BibitemShut {NoStop}%
\bibitem [{\citenamefont {Ott}\ and\ \citenamefont {Antonsen}(2008)}]{ott2008}%
  \BibitemOpen
  \bibfield  {author} {\bibinfo {author} {\bibfnamefont {E.}~\bibnamefont
  {Ott}}\ and\ \bibinfo {author} {\bibfnamefont {T.~M.}\ \bibnamefont
  {Antonsen}},\ }\href@noop {} {\bibfield  {journal} {\bibinfo  {journal}
  {Chaos: An Interdisciplinary Journal of Nonlinear Science}\ }\textbf
  {\bibinfo {volume} {18}},\ \bibinfo {pages} {037113} (\bibinfo {year}
  {2008})}\BibitemShut {NoStop}%
\bibitem [{\citenamefont {So}\ and\ \citenamefont {Barreto}(2011)}]{so2011}%
  \BibitemOpen
  \bibfield  {author} {\bibinfo {author} {\bibfnamefont {P.}~\bibnamefont
  {So}}\ and\ \bibinfo {author} {\bibfnamefont {E.}~\bibnamefont {Barreto}},\
  }\href@noop {} {\bibfield  {journal} {\bibinfo  {journal} {Chaos: An
  Interdisciplinary Journal of Nonlinear Science}\ }\textbf {\bibinfo {volume}
  {21}},\ \bibinfo {pages} {033127} (\bibinfo {year} {2011})}\BibitemShut
  {NoStop}%
\bibitem [{\citenamefont {Skardal}\ \emph {et~al.}(2014)\citenamefont
  {Skardal}, \citenamefont {Taylor},\ and\ \citenamefont
  {Restrepo}}]{skardal2014}%
  \BibitemOpen
  \bibfield  {author} {\bibinfo {author} {\bibfnamefont {P.~S.}\ \bibnamefont
  {Skardal}}, \bibinfo {author} {\bibfnamefont {D.}~\bibnamefont {Taylor}}, \
  and\ \bibinfo {author} {\bibfnamefont {J.~G.}\ \bibnamefont {Restrepo}},\
  }\href@noop {} {\bibfield  {journal} {\bibinfo  {journal} {Physica D:
  Nonlinear Phenomena}\ }\textbf {\bibinfo {volume} {267}},\ \bibinfo {pages}
  {27} (\bibinfo {year} {2014})}\BibitemShut {NoStop}%
\bibitem [{\citenamefont {Tuckwell}(1988)}]{tuckwell1988}%
  \BibitemOpen
  \bibfield  {author} {\bibinfo {author} {\bibfnamefont {H.~C.}\ \bibnamefont
  {Tuckwell}},\ }\href@noop {} {\emph {\bibinfo {title} {Introduction to
  theoretical neurobiology. Vol. 1, Linear cable theory and dendritic
  structure}}}\ (\bibinfo  {publisher} {Cambridge University Press},\ \bibinfo
  {year} {1988})\BibitemShut {NoStop}%
\bibitem [{\citenamefont {Kock}(1999)}]{kock1999}%
  \BibitemOpen
  \bibfield  {author} {\bibinfo {author} {\bibfnamefont {C.}~\bibnamefont
  {Kock}},\ }\href@noop {} {\emph {\bibinfo {title} {Biophysics of
  Computation}}}\ (\bibinfo  {publisher} {Oxford University Press},\ \bibinfo
  {year} {1999})\BibitemShut {NoStop}%
\bibitem [{\citenamefont {Izhikevich}(2000)}]{izhikevich2000}%
  \BibitemOpen
  \bibfield  {author} {\bibinfo {author} {\bibfnamefont {E.~M.}\ \bibnamefont
  {Izhikevich}},\ }\href@noop {} {\bibfield  {journal} {\bibinfo  {journal}
  {International journal of bifurcation and chaos}\ }\textbf {\bibinfo {volume}
  {10}},\ \bibinfo {pages} {1171} (\bibinfo {year} {2000})}\BibitemShut
  {NoStop}%
\bibitem [{\citenamefont {Hindmarsh}\ and\ \citenamefont
  {Rose}(1984)}]{hindmarsh1984}%
  \BibitemOpen
  \bibfield  {author} {\bibinfo {author} {\bibfnamefont {J.~L.}\ \bibnamefont
  {Hindmarsh}}\ and\ \bibinfo {author} {\bibfnamefont {R.}~\bibnamefont
  {Rose}},\ }\href@noop {} {\bibfield  {journal} {\bibinfo  {journal}
  {Proceedings of the Royal society of London. Series B. Biological sciences}\
  }\textbf {\bibinfo {volume} {221}},\ \bibinfo {pages} {87} (\bibinfo {year}
  {1984})}\BibitemShut {NoStop}%
\bibitem {suppl} See supplemental material for more details on the geometric singular perturbation analysis of the mean-
  field model.
\bibitem [{\citenamefont {Terman}(1992)}]{terman1992}%
  \BibitemOpen
  \bibfield  {author} {\bibinfo {author} {\bibfnamefont {D.}~\bibnamefont
  {Terman}},\ }\href@noop {} {\bibfield  {journal} {\bibinfo  {journal}
  {Journal of Nonlinear Science}\ }\textbf {\bibinfo {volume} {2}},\ \bibinfo
  {pages} {135} (\bibinfo {year} {1992})}\BibitemShut {NoStop}%
\bibitem [{\citenamefont {Wang}(1993)}]{wang1993}%
  \BibitemOpen
  \bibfield  {author} {\bibinfo {author} {\bibfnamefont {X.-J.}\ \bibnamefont
  {Wang}},\ }\href@noop {} {\bibfield  {journal} {\bibinfo  {journal} {Physica
  D: Nonlinear Phenomena}\ }\textbf {\bibinfo {volume} {62}},\ \bibinfo {pages}
  {263} (\bibinfo {year} {1993})}\BibitemShut {NoStop}%
\bibitem [{\citenamefont {Mosekilde}\ \emph {et~al.}(2001)\citenamefont
  {Mosekilde}, \citenamefont {Lading}, \citenamefont {Yanchuk},\ and\
  \citenamefont {Maistrenko}}]{mosekilde2001}%
  \BibitemOpen
  \bibfield  {author} {\bibinfo {author} {\bibfnamefont {E.}~\bibnamefont
  {Mosekilde}}, \bibinfo {author} {\bibfnamefont {B.}~\bibnamefont {Lading}},
  \bibinfo {author} {\bibfnamefont {S.}~\bibnamefont {Yanchuk}}, \ and\
  \bibinfo {author} {\bibfnamefont {Y.}~\bibnamefont {Maistrenko}},\
  }\href@noop {} {\bibfield  {journal} {\bibinfo  {journal} {BioSystems}\
  }\textbf {\bibinfo {volume} {63}},\ \bibinfo {pages} {3} (\bibinfo {year}
  {2001})}\BibitemShut {NoStop}%
\bibitem [{\citenamefont {Gonz{\'a}lez-Miranda}(2003)}]{gonzalez2003}%
  \BibitemOpen
  \bibfield  {author} {\bibinfo {author} {\bibfnamefont {J.~M.}\ \bibnamefont
  {Gonz{\'a}lez-Miranda}},\ }\href@noop {} {\bibfield  {journal} {\bibinfo
  {journal} {Chaos: An Interdisciplinary Journal of Nonlinear Science}\
  }\textbf {\bibinfo {volume} {13}},\ \bibinfo {pages} {845} (\bibinfo {year}
  {2003})}\BibitemShut {NoStop}%
\bibitem [{\citenamefont {Innocenti}\ \emph {et~al.}(2007)\citenamefont
  {Innocenti}, \citenamefont {Morelli}, \citenamefont {Genesio},\ and\
  \citenamefont {Torcini}}]{innocenti2007}%
  \BibitemOpen
  \bibfield  {author} {\bibinfo {author} {\bibfnamefont {G.}~\bibnamefont
  {Innocenti}}, \bibinfo {author} {\bibfnamefont {A.}~\bibnamefont {Morelli}},
  \bibinfo {author} {\bibfnamefont {R.}~\bibnamefont {Genesio}}, \ and\
  \bibinfo {author} {\bibfnamefont {A.}~\bibnamefont {Torcini}},\ }\href@noop
  {} {\bibfield  {journal} {\bibinfo  {journal} {Chaos: An Interdisciplinary
  Journal of Nonlinear Science}\ }\textbf {\bibinfo {volume} {17}},\ \bibinfo
  {pages} {043128} (\bibinfo {year} {2007})}\BibitemShut {NoStop}%
\bibitem [{\citenamefont {Coombes}\ and\ \citenamefont
  {Bressloff}(2005)}]{coombes}%
  \BibitemOpen
  \bibfield  {author} {\bibinfo {author} {\bibfnamefont {S.}~\bibnamefont
  {Coombes}}\ and\ \bibinfo {author} {\bibfnamefont {P.}~\bibnamefont
  {Bressloff}},\ }\href@noop {} {\emph {\bibinfo {title} {Bursting: the genesis
  of rhythm in the nervous system}}}\ (\bibinfo  {publisher} {World Scientific
  Publishing Company, Sigapore},\ \bibinfo {year} {2005})\BibitemShut {NoStop}%
\bibitem [{\citenamefont {Meron}(1992)}]{meron_rev}%
  \BibitemOpen
  \bibfield  {author} {\bibinfo {author} {\bibfnamefont {E.}~\bibnamefont
  {Meron}},\ }\href@noop {} {\bibfield  {journal} {\bibinfo  {journal} {Physics
  Report}\ }\textbf {\bibinfo {volume} {218}},\ \bibinfo {pages} {1} (\bibinfo
  {year} {1992})}\BibitemShut {NoStop}%
\bibitem [{\citenamefont {Ciszak}\ \emph {et~al.}(2009)\citenamefont {Ciszak},
  \citenamefont {Montina},\ and\ \citenamefont {Arecchi}}]{mchaos09}%
  \BibitemOpen
  \bibfield  {author} {\bibinfo {author} {\bibfnamefont {M.}~\bibnamefont
  {Ciszak}}, \bibinfo {author} {\bibfnamefont {A.}~\bibnamefont {Montina}}, \
  and\ \bibinfo {author} {\bibfnamefont {F.~T.}\ \bibnamefont {Arecchi}},\
  }\href@noop {} {\bibfield  {journal} {\bibinfo  {journal} {Chaos}\ }\textbf
  {\bibinfo {volume} {19}},\ \bibinfo {pages} {1} (\bibinfo {year}
  {2009})}\BibitemShut {NoStop}%
\bibitem [{\citenamefont {Dolcemascolo}\ \emph {et~al.}(2020)\citenamefont
  {Dolcemascolo}, \citenamefont {Miazek}, \citenamefont {Veltz}, \citenamefont
  {Marino},\ and\ \citenamefont {Barland}}]{dolcem}%
  \BibitemOpen
  \bibfield  {author} {\bibinfo {author} {\bibfnamefont {A.}~\bibnamefont
  {Dolcemascolo}}, \bibinfo {author} {\bibfnamefont {A.}~\bibnamefont
  {Miazek}}, \bibinfo {author} {\bibfnamefont {R.}~\bibnamefont {Veltz}},
  \bibinfo {author} {\bibfnamefont {F.}~\bibnamefont {Marino}}, \ and\ \bibinfo
  {author} {\bibfnamefont {S.}~\bibnamefont {Barland}},\ }\href@noop {}
  {\bibfield  {journal} {\bibinfo  {journal} {Physical Review E}\ }\textbf
  {\bibinfo {volume} {101}},\ \bibinfo {pages} {1} (\bibinfo {year}
  {2020})}\BibitemShut {NoStop}%
\bibitem [{SW()}]{SW}%
  \BibitemOpen
  \href@noop {} {}\bibinfo {note} {Standing waves emerge when two synchronized
  clusters of rotators oscillate with opposite angular velocity.}\BibitemShut
  {Stop}%
\bibitem [{\citenamefont {Paz{\'o}}\ and\ \citenamefont
  {Montbri{\'o}}(2009)}]{pazo2009}%
  \BibitemOpen
  \bibfield  {author} {\bibinfo {author} {\bibfnamefont {D.}~\bibnamefont
  {Paz{\'o}}}\ and\ \bibinfo {author} {\bibfnamefont {E.}~\bibnamefont
  {Montbri{\'o}}},\ }\href@noop {} {\bibfield  {journal} {\bibinfo  {journal}
  {Physical Review E}\ }\textbf {\bibinfo {volume} {80}},\ \bibinfo {pages}
  {046215} (\bibinfo {year} {2009})}\BibitemShut {NoStop}%
\bibitem [{\citenamefont {Tanaka}\ \emph {et~al.}(1997)\citenamefont {Tanaka},
  \citenamefont {Lichtenberg},\ and\ \citenamefont {Oishi}}]{tanaka1997}%
  \BibitemOpen
  \bibfield  {author} {\bibinfo {author} {\bibfnamefont {H.-A.}\ \bibnamefont
  {Tanaka}}, \bibinfo {author} {\bibfnamefont {A.~J.}\ \bibnamefont
  {Lichtenberg}}, \ and\ \bibinfo {author} {\bibfnamefont {S.}~\bibnamefont
  {Oishi}},\ }\href@noop {} {\bibfield  {journal} {\bibinfo  {journal}
  {Physical review letters}\ }\textbf {\bibinfo {volume} {78}},\ \bibinfo
  {pages} {2104} (\bibinfo {year} {1997})}\BibitemShut {NoStop}%
\bibitem [{\citenamefont {Olmi}\ and\ \citenamefont
  {Torcini}(2016)}]{olmi2016}%
  \BibitemOpen
  \bibfield  {author} {\bibinfo {author} {\bibfnamefont {S.}~\bibnamefont
  {Olmi}}\ and\ \bibinfo {author} {\bibfnamefont {A.}~\bibnamefont {Torcini}},\
  }in\ \href@noop {} {\emph {\bibinfo {booktitle} {Control of Self-Organizing
  Nonlinear Systems}}}\ (\bibinfo  {publisher} {Springer},\ \bibinfo {year}
  {2016})\ pp.\ \bibinfo {pages} {25--45}\BibitemShut {NoStop}%
\bibitem [{\citenamefont {Kuramoto}(1975)}]{kuramoto}%
  \BibitemOpen
  \bibfield  {author} {\bibinfo {author} {\bibfnamefont {Y.}~\bibnamefont
  {Kuramoto}},\ }in\ \href@noop {} {\emph {\bibinfo {booktitle} {International
  symposium on mathematical problems in theoretical physics}}}\ (\bibinfo
  {organization} {Springer},\ \bibinfo {year} {1975})\ pp.\ \bibinfo {pages}
  {420--422}\BibitemShut {NoStop}%
\bibitem [{\citenamefont {Olmi}\ \emph {et~al.}(2014)\citenamefont {Olmi},
  \citenamefont {Navas}, \citenamefont {Boccaletti},\ and\ \citenamefont
  {Torcini}}]{olmi2014}%
  \BibitemOpen
  \bibfield  {author} {\bibinfo {author} {\bibfnamefont {S.}~\bibnamefont
  {Olmi}}, \bibinfo {author} {\bibfnamefont {A.}~\bibnamefont {Navas}},
  \bibinfo {author} {\bibfnamefont {S.}~\bibnamefont {Boccaletti}}, \ and\
  \bibinfo {author} {\bibfnamefont {A.}~\bibnamefont {Torcini}},\ }\href@noop
  {} {\bibfield  {journal} {\bibinfo  {journal} {Physical Review E}\ }\textbf
  {\bibinfo {volume} {90}},\ \bibinfo {pages} {042905} (\bibinfo {year}
  {2014})}\BibitemShut {NoStop}%
\bibitem [{bim()}]{bimodal}%
  \BibitemOpen
  \href@noop {} {}\bibinfo {note} {The frequencies have been generated
  deterministically following a procedure reported in \cite{montbrio2015},
  which allows to uniformly cover the range of possible frequencies. However,
  our results are not modified by considering frequencies extracted randomly
  from bimodal Lorentzian or Gaussian distributions.}\BibitemShut {Stop}%
\bibitem [{\citenamefont {Martens}\ \emph {et~al.}(2009)\citenamefont
  {Martens}, \citenamefont {Barreto}, \citenamefont {Strogatz}, \citenamefont
  {Ott}, \citenamefont {So},\ and\ \citenamefont {Antonsen}}]{martens2009}%
  \BibitemOpen
  \bibfield  {author} {\bibinfo {author} {\bibfnamefont {E.~A.}\ \bibnamefont
  {Martens}}, \bibinfo {author} {\bibfnamefont {E.}~\bibnamefont {Barreto}},
  \bibinfo {author} {\bibfnamefont {S.~H.}\ \bibnamefont {Strogatz}}, \bibinfo
  {author} {\bibfnamefont {E.}~\bibnamefont {Ott}}, \bibinfo {author}
  {\bibfnamefont {P.}~\bibnamefont {So}}, \ and\ \bibinfo {author}
  {\bibfnamefont {T.~M.}\ \bibnamefont {Antonsen}},\ }\href@noop {} {\bibfield
  {journal} {\bibinfo  {journal} {Physical Review E}\ }\textbf {\bibinfo
  {volume} {79}},\ \bibinfo {pages} {026204} (\bibinfo {year}
  {2009})}\BibitemShut {NoStop}%
\bibitem [{\citenamefont {Pikovsky}\ and\ \citenamefont
  {Politi}(2016)}]{pikovsky2016}%
  \BibitemOpen
  \bibfield  {author} {\bibinfo {author} {\bibfnamefont {A.}~\bibnamefont
  {Pikovsky}}\ and\ \bibinfo {author} {\bibfnamefont {A.}~\bibnamefont
  {Politi}},\ }\href@noop {} {\emph {\bibinfo {title} {Lyapunov exponents: a
  tool to explore complex dynamics}}}\ (\bibinfo  {publisher} {Cambridge
  University Press},\ \bibinfo {year} {2016})\BibitemShut {NoStop}%
\bibitem [{\citenamefont {Shibata}\ and\ \citenamefont
  {Kaneko}(1998)}]{shibata1998}%
  \BibitemOpen
  \bibfield  {author} {\bibinfo {author} {\bibfnamefont {T.}~\bibnamefont
  {Shibata}}\ and\ \bibinfo {author} {\bibfnamefont {K.}~\bibnamefont
  {Kaneko}},\ }\href@noop {} {\bibfield  {journal} {\bibinfo  {journal}
  {Physical review letters}\ }\textbf {\bibinfo {volume} {81}},\ \bibinfo
  {pages} {4116} (\bibinfo {year} {1998})}\BibitemShut {NoStop}%
\bibitem [{\citenamefont {Olmi}\ \emph {et~al.}(2011)\citenamefont {Olmi},
  \citenamefont {Politi},\ and\ \citenamefont {Torcini}}]{olmi2011}%
  \BibitemOpen
  \bibfield  {author} {\bibinfo {author} {\bibfnamefont {S.}~\bibnamefont
  {Olmi}}, \bibinfo {author} {\bibfnamefont {A.}~\bibnamefont {Politi}}, \ and\
  \bibinfo {author} {\bibfnamefont {A.}~\bibnamefont {Torcini}},\ }\href@noop
  {} {\bibfield  {journal} {\bibinfo  {journal} {EPL (Europhysics Letters)}\
  }\textbf {\bibinfo {volume} {92}},\ \bibinfo {pages} {60007} (\bibinfo {year}
  {2011})}\BibitemShut {NoStop}%
\bibitem [{\citenamefont {Bick}\ \emph {et~al.}(2018)\citenamefont {Bick},
  \citenamefont {Panaggio},\ and\ \citenamefont {Martens}}]{bick2018}%
  \BibitemOpen
  \bibfield  {author} {\bibinfo {author} {\bibfnamefont {C.}~\bibnamefont
  {Bick}}, \bibinfo {author} {\bibfnamefont {M.~J.}\ \bibnamefont {Panaggio}},
  \ and\ \bibinfo {author} {\bibfnamefont {E.~A.}\ \bibnamefont {Martens}},\
  }\href@noop {} {\bibfield  {journal} {\bibinfo  {journal} {Chaos: An
  Interdisciplinary Journal of Nonlinear Science}\ }\textbf {\bibinfo {volume}
  {28}},\ \bibinfo {pages} {071102} (\bibinfo {year} {2018})}\BibitemShut
  {NoStop}%
\bibitem [{\citenamefont {Hirsch}\ \emph {et~al.}(1974)\citenamefont {Hirsch},
  \citenamefont {Devaney},\ and\ \citenamefont {Smale}}]{hirsch1974}%
  \BibitemOpen
  \bibfield  {author} {\bibinfo {author} {\bibfnamefont {M.~W.}\ \bibnamefont
  {Hirsch}}, \bibinfo {author} {\bibfnamefont {R.~L.}\ \bibnamefont {Devaney}},
  \ and\ \bibinfo {author} {\bibfnamefont {S.}~\bibnamefont {Smale}},\
  }\href@noop {} {\emph {\bibinfo {title} {Differential equations, dynamical
  systems, and linear algebra}}},\ Vol.~\bibinfo {volume} {60}\ (\bibinfo
  {publisher} {Academic press},\ \bibinfo {year} {1974})\BibitemShut {NoStop}%
\bibitem [{\citenamefont {Fenichel}(1979)}]{fenichel1979}%
  \BibitemOpen
  \bibfield  {author} {\bibinfo {author} {\bibfnamefont {N.}~\bibnamefont
  {Fenichel}},\ }\href@noop {} {\bibfield  {journal} {\bibinfo  {journal}
  {Journal of differential equations}\ }\textbf {\bibinfo {volume} {31}},\
  \bibinfo {pages} {53} (\bibinfo {year} {1979})}\BibitemShut {NoStop}%
\bibitem [{\citenamefont {Callot}\ \emph {et~al.}(1978)\citenamefont {Callot},
  \citenamefont {Diener},\ and\ \citenamefont {Diener}}]{canard1}%
  \BibitemOpen
  \bibfield  {author} {\bibinfo {author} {\bibfnamefont {J.-L.}\ \bibnamefont
  {Callot}}, \bibinfo {author} {\bibfnamefont {F.}~\bibnamefont {Diener}}, \
  and\ \bibinfo {author} {\bibfnamefont {M.}~\bibnamefont {Diener}},\
  }\href@noop {} {\bibfield  {journal} {\bibinfo  {journal} {C. R. Acad. Sci.
  Paris (Ser. I)}\ }\textbf {\bibinfo {volume} {286}},\ \bibinfo {pages} {1059}
  (\bibinfo {year} {1978})}\BibitemShut {NoStop}%
\bibitem [{\citenamefont {Benoit}\ \emph {et~al.}(1981)\citenamefont {Benoit},
  \citenamefont {Callot}, \citenamefont {Diner},\ and\ \citenamefont
  {Diener}}]{canard2}%
  \BibitemOpen
  \bibfield  {author} {\bibinfo {author} {\bibfnamefont {E.}~\bibnamefont
  {Benoit}}, \bibinfo {author} {\bibfnamefont {J.-F.}\ \bibnamefont {Callot}},
  \bibinfo {author} {\bibfnamefont {F.}~\bibnamefont {Diner}}, \ and\ \bibinfo
  {author} {\bibfnamefont {M.}~\bibnamefont {Diener}},\ }\href@noop {}
  {\bibfield  {journal} {\bibinfo  {journal} {Collect. Math.}\ }\textbf
  {\bibinfo {volume} {32}},\ \bibinfo {pages} {37} (\bibinfo {year}
  {1981})}\BibitemShut {NoStop}%
\bibitem [{\citenamefont {Marino}\ \emph {et~al.}(2007)\citenamefont {Marino},
  \citenamefont {Marin}, \citenamefont {Balle},\ and\ \citenamefont
  {Piro}}]{marino2007}%
  \BibitemOpen
  \bibfield  {author} {\bibinfo {author} {\bibfnamefont {F.}~\bibnamefont
  {Marino}}, \bibinfo {author} {\bibfnamefont {F.}~\bibnamefont {Marin}},
  \bibinfo {author} {\bibfnamefont {S.}~\bibnamefont {Balle}}, \ and\ \bibinfo
  {author} {\bibfnamefont {O.}~\bibnamefont {Piro}},\ }\href@noop {} {\bibfield
   {journal} {\bibinfo  {journal} {Physical review letters}\ }\textbf {\bibinfo
  {volume} {98}},\ \bibinfo {pages} {074104} (\bibinfo {year}
  {2007})}\BibitemShut {NoStop}%
\bibitem [{\citenamefont {Al-Naimee}\ \emph {et~al.}(2009)\citenamefont
  {Al-Naimee}, \citenamefont {Marino}, \citenamefont {Ciszak}, \citenamefont
  {Meucci},\ and\ \citenamefont {Arecchi}}]{al2009}%
  \BibitemOpen
  \bibfield  {author} {\bibinfo {author} {\bibfnamefont {K.}~\bibnamefont
  {Al-Naimee}}, \bibinfo {author} {\bibfnamefont {F.}~\bibnamefont {Marino}},
  \bibinfo {author} {\bibfnamefont {M.}~\bibnamefont {Ciszak}}, \bibinfo
  {author} {\bibfnamefont {R.}~\bibnamefont {Meucci}}, \ and\ \bibinfo {author}
  {\bibfnamefont {F.~T.}\ \bibnamefont {Arecchi}},\ }\href@noop {} {\bibfield
  {journal} {\bibinfo  {journal} {New Journal of Physics}\ }\textbf {\bibinfo
  {volume} {11}},\ \bibinfo {pages} {073022} (\bibinfo {year}
  {2009})}\BibitemShut {NoStop}%
\bibitem [{\citenamefont {Al-Naimee}\ \emph {et~al.}(2010)\citenamefont
  {Al-Naimee}, \citenamefont {Marino}, \citenamefont {Ciszak}, \citenamefont
  {Abdalah}, \citenamefont {Meucci},\ and\ \citenamefont {Arecchi}}]{al2010}%
  \BibitemOpen
  \bibfield  {author} {\bibinfo {author} {\bibfnamefont {K.}~\bibnamefont
  {Al-Naimee}}, \bibinfo {author} {\bibfnamefont {F.}~\bibnamefont {Marino}},
  \bibinfo {author} {\bibfnamefont {M.}~\bibnamefont {Ciszak}}, \bibinfo
  {author} {\bibfnamefont {S.}~\bibnamefont {Abdalah}}, \bibinfo {author}
  {\bibfnamefont {R.}~\bibnamefont {Meucci}}, \ and\ \bibinfo {author}
  {\bibfnamefont {F.}~\bibnamefont {Arecchi}},\ }\href@noop {} {\bibfield
  {journal} {\bibinfo  {journal} {The European Physical Journal D}\ }\textbf
  {\bibinfo {volume} {58}},\ \bibinfo {pages} {187} (\bibinfo {year}
  {2010})}\BibitemShut {NoStop}%
\bibitem [{\citenamefont {Popovych}\ \emph {et~al.}(2005)\citenamefont
  {Popovych}, \citenamefont {Maistrenko},\ and\ \citenamefont
  {Tass}}]{popovych2005}%
  \BibitemOpen
  \bibfield  {author} {\bibinfo {author} {\bibfnamefont {O.~V.}\ \bibnamefont
  {Popovych}}, \bibinfo {author} {\bibfnamefont {Y.~L.}\ \bibnamefont
  {Maistrenko}}, \ and\ \bibinfo {author} {\bibfnamefont {P.~A.}\ \bibnamefont
  {Tass}},\ }\href@noop {} {\bibfield  {journal} {\bibinfo  {journal} {Physical
  Review E}\ }\textbf {\bibinfo {volume} {71}},\ \bibinfo {pages} {065201}
  (\bibinfo {year} {2005})}\BibitemShut {NoStop}%
\bibitem [{\citenamefont {Tyulkina}\ \emph {et~al.}(2018)\citenamefont
  {Tyulkina}, \citenamefont {Goldobin}, \citenamefont {Klimenko},\ and\
  \citenamefont {Pikovsky}}]{tyulkina2018}%
  \BibitemOpen
  \bibfield  {author} {\bibinfo {author} {\bibfnamefont {I.~V.}\ \bibnamefont
  {Tyulkina}}, \bibinfo {author} {\bibfnamefont {D.~S.}\ \bibnamefont
  {Goldobin}}, \bibinfo {author} {\bibfnamefont {L.~S.}\ \bibnamefont
  {Klimenko}}, \ and\ \bibinfo {author} {\bibfnamefont {A.}~\bibnamefont
  {Pikovsky}},\ }\href@noop {} {\bibfield  {journal} {\bibinfo  {journal}
  {Physical review letters}\ }\textbf {\bibinfo {volume} {120}},\ \bibinfo
  {pages} {264101} (\bibinfo {year} {2018})}\BibitemShut {NoStop}%
\bibitem [{\citenamefont {Goldobin}\ and\ \citenamefont
  {Klimenko}(2020)}]{goldobin2020}%
  \BibitemOpen
  \bibfield  {author} {\bibinfo {author} {\bibfnamefont {D.~S.}\ \bibnamefont
  {Goldobin}}\ and\ \bibinfo {author} {\bibfnamefont {L.~S.}\ \bibnamefont
  {Klimenko}},\ }in\ \href@noop {} {\emph {\bibinfo {booktitle} {AIP Conference
  Proceedings}}},\ Vol.\ \bibinfo {volume} {2216}\ (\bibinfo {organization}
  {AIP Publishing LLC},\ \bibinfo {year} {2020})\ p.\ \bibinfo {pages}
  {070001}\BibitemShut {NoStop}%
\bibitem [{\citenamefont {Montbri{\'o}}\ \emph {et~al.}(2015)\citenamefont
  {Montbri{\'o}}, \citenamefont {Paz{\'o}},\ and\ \citenamefont
  {Roxin}}]{montbrio2015}%
  \BibitemOpen
  \bibfield  {author} {\bibinfo {author} {\bibfnamefont {E.}~\bibnamefont
  {Montbri{\'o}}}, \bibinfo {author} {\bibfnamefont {D.}~\bibnamefont
  {Paz{\'o}}}, \ and\ \bibinfo {author} {\bibfnamefont {A.}~\bibnamefont
  {Roxin}},\ }\href@noop {} {\bibfield  {journal} {\bibinfo  {journal}
  {Physical Review X}\ }\textbf {\bibinfo {volume} {5}},\ \bibinfo {pages}
  {021028} (\bibinfo {year} {2015})}\BibitemShut {NoStop}%
\end{thebibliography}

%

\end{document}